\documentclass[prb, preprint, 12pt, groupedaddress, showpacs]{revtex4}
\usepackage{amsmath, graphicx, epsfig}
\begin{document}
\title{Is charge noise in single electron transistors and charge qubits caused by metallic grains?}
\author{S. Kafanov$^1$}
\email{sergey.kafanov@mc2.chalmers.se}
\author{H. Brenning$^1$}
\author{T. Duty$^{1,2}$}
\author{P. Delsing$^1$}
\affiliation{$^1$Microtechnology and Nanoscience, Chalmers University of
Technology; S-41296, G\"{o}teborg, Sweden.}
\affiliation{$^2$School of Physical Sciences, University of Queensland;
St. Lucia, QLD 4072 Australia.}

\date{\today}


\begin{abstract}
We report on measurements of low frequency noise in a single electron transistor from a few Hz up to $10\,\mathrm{MHz}$. Measurements were done for different bias and gate voltage, which allows us to separate noise contributions from different noise sources. We find a 1/f noise spectrum with two Lorentzians superimposed. The cut-off frequency of one of the Lorentzians varies systematically with the potential of the SET island. Our data is consistent with two single-charge fluctuators situated close to the tunnel barrier. We suggest that these are due to random charging of aluminum grains, each acting as a single electron box with tunnel coupling to one of the leads and capacitively coupled to the SET island. We are able to fit the data to our model and extract parameters for the fluctuators.
\end{abstract}

\maketitle

\section{Introduction}{\label{Introduction}}
Single electron transistors (SETs)
\cite{IEEE.Transactions.on.Magnetics.1987.1142-5,
PhysRevLett.59.109} are extremely sensitive electrometers, with demonstrated charge sensitivities of the order of $\mathrm{\mu e/\sqrt{Hz}}$ \cite{ApplPhysLett.79.4031, JApplPhys.100.114321}. Due to their high charge sensitivities they have found a large number of applications in research, for example, SETs are used to detect nano-mechanical oscillators \cite{Science.304.74}, to count electrons \cite{Nature.423.422, Nature.434.361} and to read out qubits \cite{PhysRevB.69.140503, PhysRevLett.90.027002,
Nature.421.823}.

The fundamental limitation for the sensitivity of the SET is set by shot noise generated when electrons tunnel across the tunnel barriers \cite{Korotkov, Nature.406.1039}. Shot noise was observed in a two junction structure (without gate) by Birk \emph{et al.} \cite{PhysRevLett.75.1610}.

However, there are two other types of noise which are limiting the charge sensitivity in experiments. At high frequencies, above $1\,\mathrm{MHz}$, the sensitivity is limited by the amplifier noise. At low frequencies the sensitivity is limited by $1/f$ noise which is due to background charge fluctuators near the SET. A collection of several fluctuators with different frequencies leads to a $1/f$ spectrum. In several cases it has been observed that there is a background of $1/f$ noise and a single or a few more strongly coupled fluctuators, resulting in random telegraph noise, which in the frequency spectrum leads to Lorentzians superimposed on the $1/f$ background \cite{ApplPhysLett.61.237, ApplSupercondIEEETrans.5.3085}.

Understanding the nature of the $1/f$ noise is also very important for solid state qubits since $1/f$ noise strongly limits the decoherence time for these qubits. It has been suggested by Ithier \emph{et al.} that the charge noise has a cut-off  at the frequency of the order of $0.5\,\mathrm{MHz}$ \cite{PhysRevB.72.134519}.

Even though there have  been many efforts \cite{ApplPhysLett.61.237,
ApplSupercondIEEETrans.5.3085, IEEETransInstrMeas.46.303,
JApplPhys.84.3212, JApplPhys.78.2830, Bouchiat, PhysRevB.56.7675,
JApplPhys.86.2132, JApplPhys.83.310, JLowTempPhys.123.103} to reveal the physical origin of the background charge fluctuators the nature of these fluctuators is still unknown. It is not even clear where these fluctuators are located. The charge fluctuators can be located either in the tunnel barrier or outside the barrier but in close proximity of the junction.

The role of the substrate has been examined in several experiments
\cite{Bouchiat, JApplPhys.84.3212, ApplPhysLett.91.033107}. However, those experiments did not show a strong dependence of the noise on the substrate material. The barrier dielectric has been proposed as location of the charge traps by several groups \cite{ApplSupercondIEEETrans.5.3085, JApplPhys.84.3212,
JApplPhys.78.2830, PhysRevB.53.13682}.

Several groups have shown that the low frequency noise at the output of the SET varies with the current gain ($\partial I/\partial Q_g$) of the SET and that the maximum noise is found at the bias point with maximum gain \cite{JApplPhys.78.2830, JApplPhys.86.2132, JApplPhys.83.310}. This indicates that the noise sources acts at the input of the device, \emph{i.e.} as an external fluctuating charge. A detailed comparison of the noise to the gain was done by Starmark \emph{et al.} \cite{JApplPhys.86.2132}. All above mentioned experiments were performed with conventional SETs by measuring current or voltage noise at relatively low frequencies. \emph{i.e.} below a few kHz.

In this work we have measured low frequency noise in the Single Electron Transistor which has demonstrated a very high charge sensitivity
\cite{JApplPhys.100.114321}, by using the Radio Frequency Single Electron Transistor (rf-SET) technique \cite{Science280.1238, Wahlgren}. This allowed us to measure low frequency noise of the reflected voltage from the rf-SET in the range of a few hertz up to tens of MHz, and due to high charge sensitivity we were not limited by the amplifier noise. We find two Lorentzians superimposed on a $1/f$ spectrum and that the noise in the range$50\,\mathrm{kHz}$-$1\,\mathrm{MHz}$ is quite different for positive and for negative gain of the transistor. By analyzing the bias and gate dependence of the noise we argue that the noise in this frequency range is dominated by electron tunneling to an aluminum grain, which acts as a single electron box capacitively connected to the SET island.

The paper is organized as follows. In section \ref{LFN_model} we describe a model for the low frequency noise, which allows us to separate contributions from the different noise sources. In section \ref{Experimental_detail} we describe the experimental details of our measurements. Section \ref{Experimental_results} is the main part of this paper and contains the experimental results. Finally in section \ref{Discussion} we describe our model for the nature of the low frequency noise in our SETs.

\section{Low-Frequency Noise Model}{\label{LFN_model}}
We start by analyzing the different contributions of the measured noise. What we actually measure is the reflected voltage from the tank circuit in which the SET is embedded. The rf-SET tank circuit is a series $LC$ circuit with inductance $L$ and capacitance $C$. The power spectral density of the reflected voltage can be decomposed into the following terms originating from charge noise,
resistance noise, shot noise and amplifier noise according to the following equation \cite{Korotkov_unpubl, JApplPhys.86.2132}:
\begin{equation}{\label{Noise_decomp}}
S_{|v_{r}|}=\left(\frac{\partial |v_r|}{\partial
Q_g}\right)^2S_{Q_g}(f)+ \left(\frac{\partial |v_r|}{\partial
R_1}\right)^2S_{R_{1}}(f)+\left(\frac{\partial |v_r|}{\partial
R_2}\right)^2S_{R_{2}}(f)+S_{Shot}+S_{Ampl.},
\end{equation}
where $Q_g$ is the charge at the input gate and $R_{1,2}$ are the tunnel resistances of the two junctions. Here we neglected the higher order terms and possible correlation terms between the charge noise and the resistance noise. In the case when the charge fluctuator is located in the tunnel barrier the correlation term may not be negligible. By measuring the noise for different bias points having different gains (\emph{i.e.} different $\partial |v_r|/\partial Q_g$) it is possible to extract information on whether the noise is associated with charge or resistance fluctuators.

We have designed the matching circuit for the rf-SET to work in the over-coupled regime, in order to have a monotonic dependence of the reflection coefficient as a function of the SET differential conductance. This regime corresponds to the matching condition when the internal quality factor of the rf-SET tank circuit
($Q_{int}=\sqrt{L/C}/R$) is larger than the external quality factor 
($Q_{ext}=Z_0/\sqrt{L/C}$), where $Z_0=50\,\Omega$ is the characteristic impedance of the transmission line, and $R$ is the SET differential resistance.

We have theoretically analyzed the reflected voltage from the rf-SET as a function bias and gate voltages applied to the SET. The reflected voltage characteristic of the rf-SET can be calculated by the orthodox theory \cite{Averin} using a master equation approach. From this theory we can also calculate the derivatives in eq.(\ref{Noise_decomp}) of the reflected voltage with respect to variations in gate charge and in the resistances of the SET tunnel junctions.

\begin{figure}
\centering\epsfig{figure=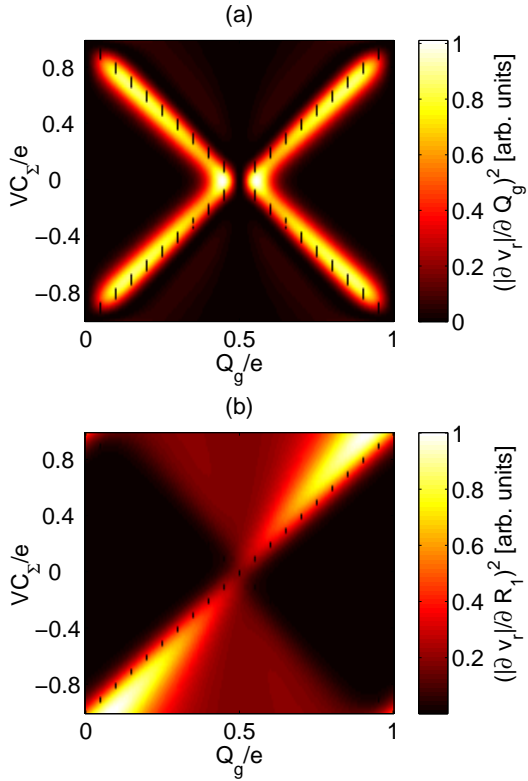,width=7cm}
\caption{\label{Orthodox}\small(color online) Calculated derivatives from equation (\ref{Noise_decomp}) as a function of bias voltage $V$, and gate charge $Q_g=C_gV_g$. The derivative of the reflected voltage from the rf-SET with respect to (a) the charge fluctuations $(\partial |v_r|/\partial Q_g)^2$; (b) the resistance fluctuations in the 1st junction $(\partial |v_r|/\partial R_1)^2$. Sensitivity to the resistance fluctuations in the second junction has a mirror symmetry, along SET open state $(Q_g=0.5\,e)$, with the sensitivity to the resistance fluctuations in the first barrier.}
\end{figure}

Figure \ref{Orthodox}(a) shows the sensitivity of the rf-SET to charge fluctuations as a function of the bias and gate voltages. The charge sensitivity is a symmetric function around the SET open state ($Q_g=0.5\,e$), and has maxima close to the onset of the open state.

The sensitivity of the SET to resistance fluctuations in the first tunnel barrier is shown in the figure \ref{Orthodox}(b). The sensitivity to resistance fluctuations in the second barrier (not shown) is identical to figure \ref{Orthodox}(b), except that it is mirrored along the SET open state $(Q_g=0.5\,e)$.

By operating at different bias and gate voltage we can choose operation points where the noise contribution from the different derivatives dominates, and it is possible to distinguish charge noise from resistance noise.

\section{Experiment details}
{\label{Experimental_detail}}
The samples were fabricated on oxidized silicon substrates using electron beam lithography and a standard double-angle evaporation technique. The asymptotic resistance of the measured single electron transistor was $25\,\mathrm{k\Omega}$. The charging energy $E_C=e^2/2C_\Sigma\simeq 18\pm 2\,\mathrm{K}$ was extracted from the measurements of the SET stability diagram of the reflected signal with frequency $f=350\,\mathrm{MHz}$ versus bias voltage and gate voltages. From the asymmetry of the SET stability diagram we could
also deduce that the asymmetry in the junction capacitances was $30\%$.

The SET was embedded in a resonant circuit and operated in the radio frequency mode \cite{Science280.1238, Wahlgren}. The bandwidth of the setup was approximately $10\,\mathrm{MHz}$ limited by the quality factor of the resonance circuit. The radio frequency signal was launched toward the resonance circuit and the reflected signal was amplified by two cold amplifiers, and then downconverted using homodyne mixing. The output signal from the mixer containing the noise information was then measured by a spectrum analyzer. The sample was attached to the mixing chamber of a dilution refrigerator which was cooled to a temperature of approximately $25\,\mathrm{mK}$. All measurements were performed in the normal (nonsuperconducting) state at a magnetic field of $1.5\,\mathrm{T}$.

We have performed the noise measurements for different gate voltages and different bias voltages. Due to experimental problems we have performed measurements mostly for negative biases. The sample shows very high charge sensitivity of the order of $1\,\mathrm{\mu e/\sqrt{Hz}}$. For a detailed description of the sensitivity with respect to different parameters (see
ref.\cite{JApplPhys.100.114321}). A small sinusoidal charge signal
of $7.3\cdot 10^{-4}\,\mathrm{e_{rms}}$ with a frequency of $133\,\mathrm{Hz}$ was applied to the gate, which allowed us to calibrate the charge sensitivity referred to the input of the SET.

\begin{figure}
\centering\epsfig{figure=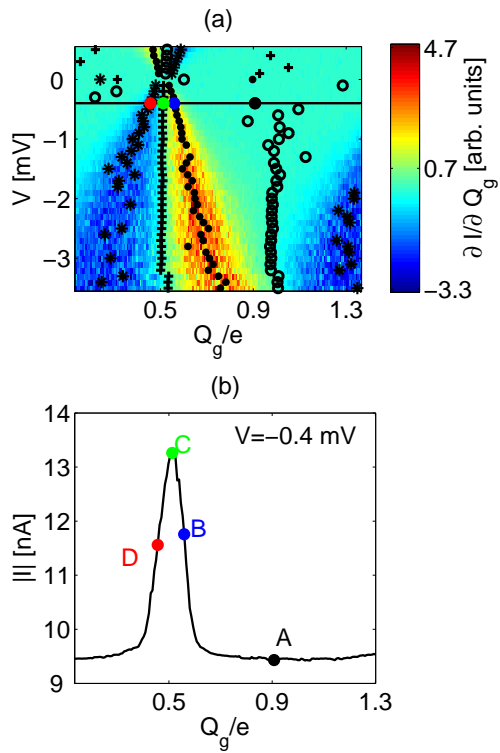,width=7cm}
\caption{\label{Points}\small(color online) (a) The SET stability diagram ($\partial I/\partial Q_g$) measured at a temperature of $25\,\mathrm{mK}$. Horizontal black line corresponds to the bias voltage, where transfer function $I(Q_g)$ (see panel (b)) was obtained. The points \textbf{A}/ (($\circ$) for negative and ($+$) for positive bias) where $I=0$ inside the Coulomb blockade region. The points \textbf{B} and \textbf{D} ($\ast/\bullet$) correspond to
maximum positive/negative gain, $\partial I/\partial Q_g$ that is where $\partial^2I/\partial Q_g^2=0$. The measurement points \textbf{C}, close to the current maximum current, correspond to $\partial I/\partial Q_g=0$, marked with ($+$) at negative bias and with ($\circ$) at positive bias. (b) SET transfer function $I(Q_g)$ measured for the bias voltage $V=-0.4\,\mathrm{mV}$. In the stability diagram this measurements is shown as a black line.}
\end{figure}

Before each measured spectrum, we have employed a charge-locked loop
\cite{ApplPhysLett.81.4859}, with a help of a lock-in amplifier. The first ($\partial I/\partial Q_g$) or the second ($\partial^2 I/\partial Q_g^2$) derivatives of the current were used as an error signals for stabilization of the gate point to compensate for the slow drift at the current maximum points (\textbf{C} in fig.\ref{Points}(b)), or at the maximum gain points (\textbf{B} and \textbf{D} in fig.\ref{Points}(b)) respectively. Each noise spectrum
is however measured with the feed back loop turned off.

\section{Experimental Results}
{\label{Experimental_results}}
In order to separate the contributions from different noise sources we have performed measurements at specific points. The measurement points are shown in fig. \ref{Points}. At point \textbf{A} ($Q_g\approx 0\,e$) there is an almost a complete Coulomb blockade with a zero current and zero gain ($\partial |v_r|/\partial Q_g=0$). Here the derivatives of the reflected voltage with respect to resistance fluctuations in the tunnel barriers ($\partial
|v_r|/\partial R_{1,2}=0$) are also zero (see fig.\ref{Orthodox}(b)).
At this point we see only amplifier noise --- different curves for different bias voltages show the same noise level convincing us that we see only amplifier noise. These measurements thus serve to calibrate the noise of the amplifier.

The measurements at points \textbf{B} and \textbf{D}, correspond to the requirement of maximum current gain ($\max|\partial I/\partial Q_g|$) and therefore also high $|\partial |v_r|/\partial Q_g|$, diagonals in figure\ref{Orthodox}(a). At these points there are contributions from all the noise sources (see fig. \ref{Orthodox}), but since the absolute gain and the current are very similar at the points \textbf{B} and \textbf{D}, these measurements can be compared.

We have also measured noise at the points \textbf{C} ($Q_g\approx 0.5\,e$) close to the maximum of the current transfer function ($|I(Q_g)|$). At this point the gain of the reflected signal ($\partial |v_r|/\partial Q_g$) is quite low, but the shot noise could be high. The derivatives of the reflected voltage with respect to resistance fluctuations in the tunnel barriers ($\partial |v_r|/\partial R_{1,2}$) are small but not equal to zero (see fig. \ref{Orthodox}(b)).
\begin{figure}
\centering\epsfig{figure=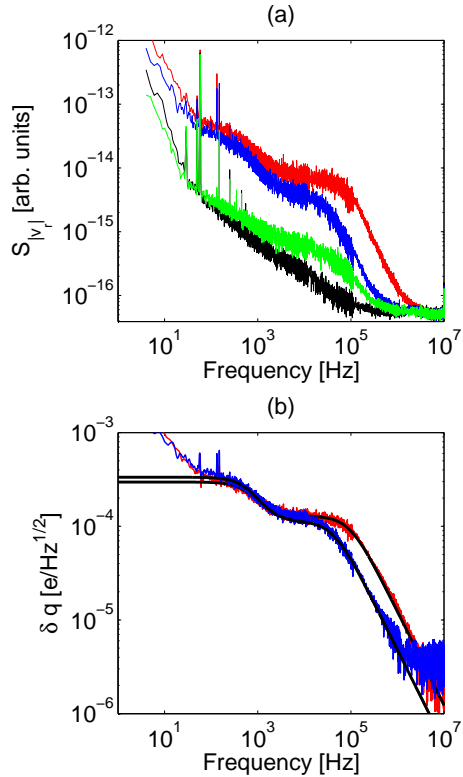,width=7cm}
\caption{\label{Noise}\small (color online) (a) Power spectral density (PSD) of the reflected voltage measured in the points \textbf{A} (black curve); \textbf{B} (blue curve); \textbf{C} (green curve); \textbf{D} (red curve). (b) Normalized noise in the points \textbf{B} (blue curve); \textbf{D} (red curve). The black continuous lines are a fits to the measured PSD with a sum of two Lorentzian functions.}
\end{figure}

The noise of the reflected voltage at the fixed bias point and for the different gate points are shown in fig. \ref{Noise}(a). We start by subtracting the amplifier noise and then we compare the noise measured at the different points. Comparing the noise measured at points \textbf{B} and \textbf{D}, where the current gain has a maximum, with the noise measured at point \textbf{C} close to the maximum of the transfer function we see that the noise at the point
\textbf{C} is substantially lower, even though the current is higher. From this we draw the conclusion that the difference is not due to the shot noise.

Comparing the noise spectra measured at the points \textbf{B} and \textbf{D} it is necessary to note that both the currents and the gains are very similar at these points. Both spectra \textbf{B} and \textbf{D} have a $1/f$ dependence at low frequencies with two Lorentzian shoulders at higher frequencies.  At frequencies above $30\,\mathrm{kHz}$ the noise at point \textbf{D} drops well bellow the noise at point \textbf{B}. At $300\,\mathrm{kHz}$ the difference is a factor of 5.

We have fitted the noise spectra at points \textbf{B} and \textbf{D} to a sum of two Lorentzian functions. The results of these fits are shown in figure \ref{Noise}(b). From these fits we can extract the cut-off  frequency and the level for each of the two Lorentzians.

The low-frequency Lorentzian has a cut-off frequency of the order of $1\,\mathrm{kHz}$. The cut-off frequency is the same for both slopes ($\partial I/\partial Q_g \,^>_< 0$), and it does not show any bias or gate dependence within the accuracy of our measurements.

\begin{figure}[b]
\centering\epsfig{figure=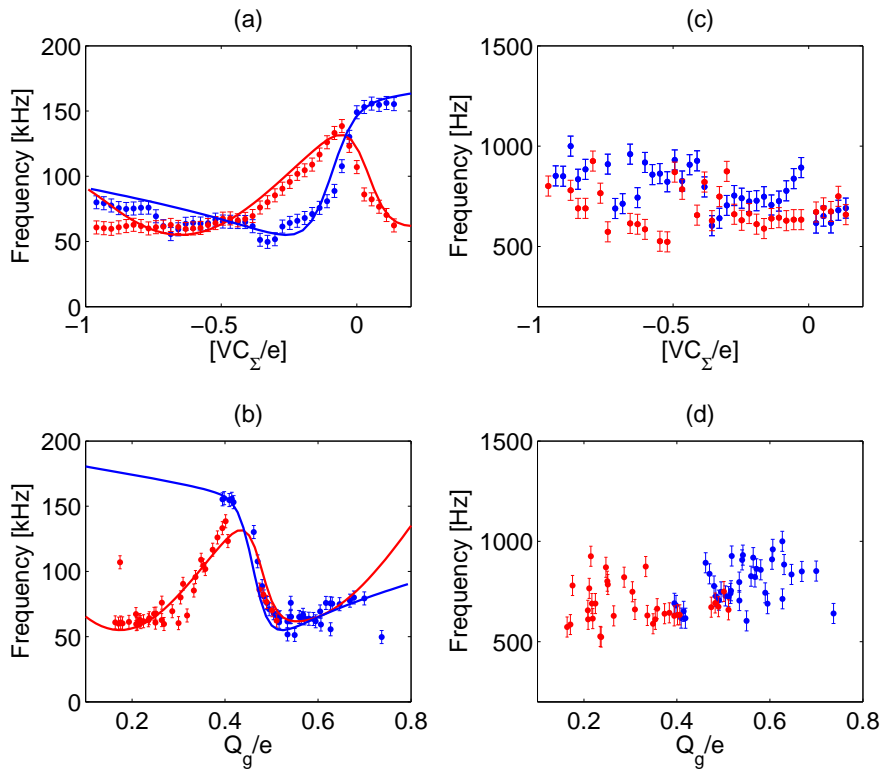,width=12cm}
\caption{\label{Gate_Depend}\small (color online) (a) The bias dependence of the cut-off frequency  for the high frequency Lorentzian. (b) The gate dependence of the cut-off frequency for the high frequency Lorentzian. (c) The bias dependence of the cut-off frequency  for the low frequency Lorentzian. (d) The gate dependence of the cut-off frequency for the low frequency Lorentzian. Blue points correspond to the negative slope $\partial I/\partial Q_g>0$ (see fig. \ref{Points}). Red points correspond to the positive slope $\partial I/\partial Q_g<0$ (see fig.  \ref{Points}). The error bars are extracted from the fits to the Lorentzians. The continuous lines (blue, red) show the bias and gate dependencies for the Lorentzian cut-off frequency calculated in the described model.}
\end{figure}

In contrast, the high-frequency Lorentzian has a cut-off frequency above $f_{co}>50\rm\,kHz$, with a strong dependence on the bias and gate voltages. The bias dependence of the Lorentzian cut-off frequency for the positive ($\partial I/\partial Q_g>0$) and negative ($\partial I/\partial Q_g <0$) slopes are shown in figure \ref{Gate_Depend}(a).

For the negative slope $(\partial I/\partial Q_g<0)$(blue points in figure \ref{Gate_Depend}(a)), the cut-off frequency remains practically constant ($f_{co}\simeq 50\rm\, kHz$) for negative biases. Close to the zero bias voltage the Lorentzian cut-off frequency switches to a higher frequency ($f_{co}\simeq 150\rm\, kHz$). For the positive slope $(\partial I/\partial Q_g>0)$ (red curve in figure \ref{Gate_Depend}), the situation is different. For negative bias voltage the Lorentzian cut-off frequency is continuously growing from $f_{co}\simeq 60\rm\, kHz$ and reaches a maximum ($f_{co}\simeq 120\rm\, kHz$) close to zero bias voltage. For the positive bias voltage it rapidly decrease from the maximum to the initial value $f_{co}\simeq 60\rm\, kHz$).
Figure \ref{Gate_Depend}(b), shows the gate dependence for the Lorentzian cut-off frequencies for both slopes $(\partial I/\partial Q_g\,^>_<0)$. As is clearly shown in this figure, the gate dependence for the positive and negative slopes are different. The gate dependence for the positive slope $(\partial I/\partial Q_g>0)$ (red curve) has a peak, with a small negative offset on the gate charge from the SET open state. The Lorentzian cut-off frequency behavior on the negative slope $(\partial I/\partial Q_g<0)$ (blue curve) is a step like function.

By integrating the Lorentzians in the noise spectra (see figure \ref{Noise}(b)) we can calculate the total variation of induced charge on the SET island for both fluctuators. The variation of induced charge, for the low-frequency fluctuator is of the order of $\Delta q_{lf}\approx 6.6\,\mathrm{me_{rms}}$. The same estimation for the high-frequency fluctuator gives $\Delta q_{hf}\approx 30\,\mathrm{me_{rms}}$.

\section{Discussion}{\label{Discussion}}
In this section we analyze the bias and the gate dependence of the cut-off frequency of the high frequency Lorentzian ($f_{co}\simeq 50-150\,\mathrm{kHz}$). In particular, we try to explain why the cut-off frequency is different for different biasing conditions.

In our analysis we have assumed that there are in principle two possible sources for the low frequency noise, resistance fluctuators or charge fluctuators. The physical nature of the resistance fluctuators is not well understood, but they can be related with charge fluctuations. For instance, a charge oscillating in the tunnel barrier may modify both the transparency and the induced island charge. 

The noise from a resistance fluctuator in one of the tunnel barriers would have an asymmetry along the onset of the SET open state, as it was shown in section \ref{LFN_model} (see fig. \ref{Orthodox}(b)). In order to explain the bias dependence of the cut-off frequency (see fig. \ref{Gate_Depend}(a)) in terms of resistance fluctuators we must assume that there is an individual resistance fluctuator located in each of the SET tunnel barriers. Furthermore these fluctuators must have the same tunneling rates. With this strong assumption, however, it is impossible to explain the sharp drop of the experimentally measured gate dependence of the cut-off frequency (see. fig. \ref{Gate_Depend}(b)).

Thus, in order to explain the results for the high-frequency Lorentzian, we will assume that there are individual charge fluctuators affecting the SET, and that each Lorentzian in the experimentally measured spectra is due to a single fluctuator coupled to the SET island. Many experimental groups have suggested a microscopic nature of these fluctuators. The microscopic nature is not known well, but there are suggestions that it could be traps in the substrate dielectric close to the SET or in the aluminum surface oxide. 

Here we will argue that the sources of these two level fluctuators are located outside the barrier and that they may have a \textit{mesoscopic} nature. In reference \cite{PhysRevB.53.13682} it is argued based on electrostatic analysis of the tunnel barrier, that such fluctuators could not be localized inside barrier. There are also other experiments \cite{PhysRevB.67.205313, JApplPhys.96.6827}, where it is argued that the charge fluctuators most probably are localized outside the tunnel barrier.

\begin{figure}
\centering\epsfig{figure=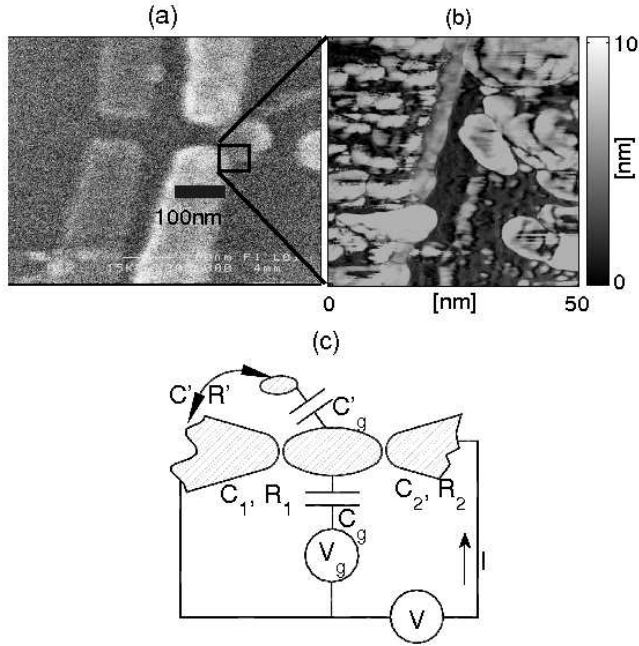,width=10cm}
\caption{{\label{SET_SEB}}\small (a) A SEM image of a sister sample to the measured one. The black bar shows $100\,\mathrm{nm}$ linear scale. (b) An AFM pictures of the edge of an aluminum film on the $SiOx$ surface (c) Equivalent electrostatic scheme, where the small metallic grain is capacitively coupled to the SET island and has a tunnel connection with a bias lead.}
\end{figure}

A typical SET is made from thin aluminum films which are not uniform; they consist of small grains connected to each other. In figure \ref{SET_SEB}(a) we are show a SEM image of a sister sample to the measured one. In figure \ref{SET_SEB}(b) we also show an AFM image of the same sister sample. It can be clearly seen that there are many small grains close to the device. We will assume that some grains are separated from the main film by a thin oxide layer but also capacitively connected to the SET island.

Electrostatically such a grain can be described as a single electron box \cite{ZeitsPhysB.85.327}, capacitively coupled to the SET island with capacitance $C'_g$ and having a tunnel contact with resistance $R'$ and capacitance $C'$ with one of the bias leads as indicated in figure\,\ref{SET_SEB}(c). The situation is almost equivalent if the grain would be tunnel connected to the SET island and capacitively connected to the SET bias lead. For a detailed analysis we should estimate the energy scales for this grain. We assume that the linear dimension of the stray aluminum grain is in the range $1-5\,\mathrm{nm}$. The charging energy for this grain is of the order $E'_C \equiv e^2/(2 C'_\Sigma) \sim 10^{-1}\,\mathrm{eV}$, where $C'_\Sigma=C'+C'_g+C'_{env}$, and  $C'_{env}$ is the capacitance to the rest of the environment. This charging energy is substantially larger than the experimental temperature and the charging energy of the SET ($k_\mathrm{B}T \ll E_C \ll E'_C$). In addition there will be be further separation of the energy levels due to the small size of the grains. 

The ratio of capacitances $C'_g/C'_\Sigma$ is given directly by the charge induced on the SET island which we already have extracted by integrating the Lorentzians. Thus for the high frequency Lorentzian we have $C'_g/C'_\Sigma=0.030$ and for the low frequency Lorentzian this ratio is 0.0066.

In our model the single electron box is capacitively coupled to the SET island, and tunnel coupled with one of the SET leads. The average potential of the SET island $\phi$ acts, in this system, as a gate potential  for the single electron box and induces the charge $n'_g=C'_g \phi /e$ on the grain.
The charging dynamics for the electron box can be described by the orthodox theory using a master equation approach \cite{Averin}. Electron tunneling changes the number of excess electrons $n$ in the grain. The differences in the electrostatic energy, when electrons tunnel to $(+)$ and from $(-)$ the grain are:
\begin{equation}
\Delta\mathcal{E}_c^{\pm}(n)=2 E'_C \left( \pm n \mp n'_g +1/2\right).
\end{equation}

The tunnel rates of electron tunneling to or from the grain is a function of the tunnel resistance $R'$ and the electrostatic energy gain $\Delta \mathcal{E}_c^{\pm}(n)$:
\begin{equation}{\label{Tunnel_rates}}
w^{\pm}_n=\frac{1}{e^2 R'}
\frac{\Delta \mathcal{E}_c^{\pm}(n)}{1-\exp\left(-\Delta \mathcal{E}_c^{\pm}(n)/(k_\mathrm{B}T)\right)},
\end{equation}

The probability $\sigma_n$ to have $n$ excess electrons in the grain
obeys the master equation:
\begin{equation}
\frac{d\sigma_n}{dt}=w^+_{n-1}\sigma_{n-1}+w^-_{n+1}\sigma_{n+1}- \sigma_n(w^+_n+w^-_n)
\end{equation}

For our case of low temperature, $k_B T \ll E'_C$ 
there are only two nonvanishing probabilities, $\sigma_n$ and $\sigma_{n+1}$.  This simplifies the problem, and it is convenient to define the distance from the grain degeneracy point, \textit{i.e.} the point where these two nonvanishing states are degenerate, as $\Delta n'_g=C'_g (\phi -\phi_0) /e$. Here $\phi_0=e(n+1/2)/C'_g$ is the SET island potential needed to reach the grain degeneracy point. 

In the time domain, the dynamics for charging this grain from the lead is a random telegraph process, and the spectrum of this process is a Lorentzian function with a cut-off frequency defined by the sum of charging and escaping rates \cite{JApplPhys.25.341}:
\begin{equation}{\label{Eq_Cut_off frequency}}
f_{co} = w^{+}_{n}+w^{-}_{n+1} = \frac { \Delta n'_g}{R' C'_\Sigma} \coth \left( \frac{E'_C}{k_B T} \Delta n'_g \right)
= \mathcal{A}(\phi -\phi_0) \coth \left( \mathcal{B} (\phi -\phi_0) \right),
\end{equation}
where $\mathcal{A}=C'_g/(e C'_\Sigma R')$ and $\mathcal{B}=e C'_g/(2 k_B T C'_\Sigma)$.

From eq. (\ref{Eq_Cut_off frequency}) we thus see that the cut-off frequency of the Lorentzian depends directly on $\Delta n'_g$ and therefore on the potential of the SET island, $\phi$. When the grain is biased away from its degeneracy point, \textit{i.e.} when $\Delta n'_g > 2k_B T/E'_C$, the cut-off frequency grows linearly with the SET island potential. This means that if we are far from the grain degeneracy point, the cut-off frequency is relatively high and the relative frequency shift due to the change in $\phi$ will be small. On the other hand if we are close to the grain degeneracy point the cut-off frequency will be close to its minimum and the relative change in frequency due to $\phi$ can be substantial. The maximum relative frequency change occurs when the potential just barely reaches the grain degeneracy point and can be calculated from eq.\,\ref{Eq_Cut_off frequency}. Taking in to account the bounds of the island potential, $-e/(2C_\Sigma)<\phi<e/(2C_\Sigma)$, we get
\begin{equation}{\label{freq_change}}
\left.\frac{\Delta f_{co}}{f_{co,min}}\right|_{max} =  \mathcal{B}\Delta\phi_{max} \coth \left( \mathcal{B} \Delta \phi_{max} \right) -1= 
\frac{C'_g}{C'_\Sigma}\frac{E_C}{k_B T}\coth \left( \frac{C'_g}{C'_\Sigma}\frac{E_C}{k_B T} \right)-1
\end{equation}

We note that the relative frequency shift is independent of $R'$ and that a large charging energy of the SET is important to observe a frequency shift. Thus it is clear that the relative frequency shift of most Lorentzians will be very small. To observe a frequency shift as large as in figure\,\ref{Gate_Depend}, the the grain will have to be close to its degeneracy point, and in addition the charging energy of the SET will have to be large, to create a substantial frequency shift. Obviously a relatively strong coupling between the grain and the SET island is also important.

\begin{figure}
\centering\epsfig{figure=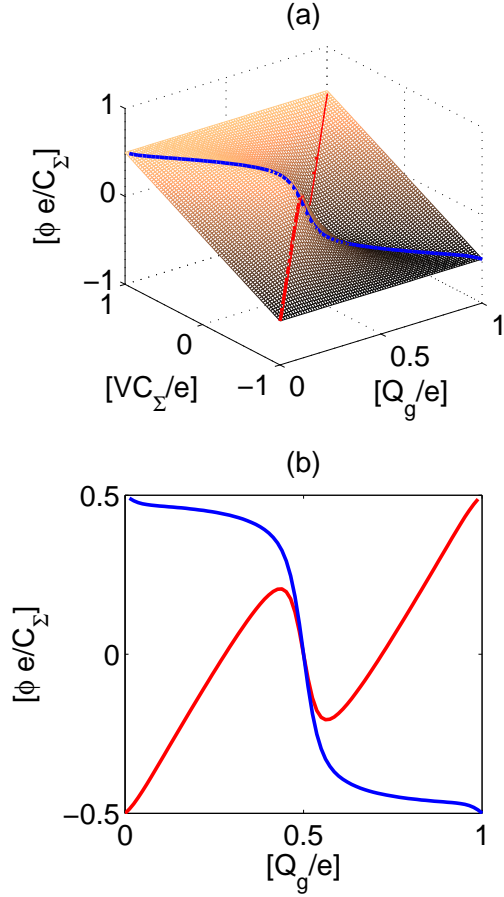,width=7cm}
\caption{{\label{Potential}}\small (color online) (a) The SET island potential $\phi$ calculated from the orthodox theory, as a function of bias voltage and gate charge. (b) Line cuts along the two diagonals in (a), where the measurements have been performed. As can be seen the potential is asymmetric with respect to the SET degeneracy point. If the grain SEB is biased away from its degeneracy point the frequency of the tunneling on and off the grain should be proportional to the island potential. The resulting fit is shown in figure\,\ref{Fit}.}
\end{figure}

The average potential of the SET island will depend both on the bias and gate voltages of the SET, $\phi(V,V_g)$, and can be calculated using the orthodox theory \cite{Averin}. We have calculated the potential as a function of bias voltage and gate voltage, as is shown in figure\,\ref{Potential}(a), and in particular we have calculated the potential along the two diagonals (roughly where the measurements have been performed) which is shown in figure\,\ref{Potential}(b) to give an idea of how the potential varies (\textit{c.f.} figures\,\ref{Orthodox} and \ref{Points}(a)). As can be seen, the potential is asymmetric with respect to the SET degeneracy point and varies between $-e/(2C_\Sigma)$ and $e/(2C_\Sigma)$.  

\begin{figure}
\centering\epsfig{figure=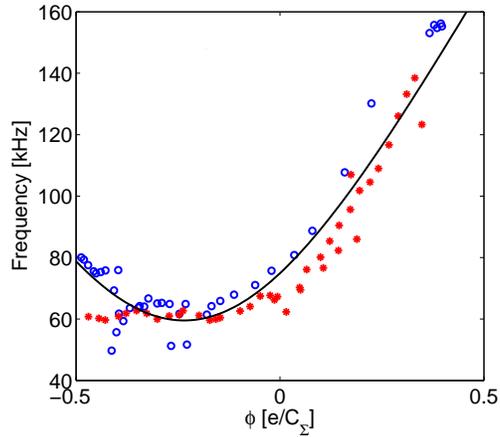,width=7cm}
\caption{{\label{Fit}}\small (color online) The cut off frequency of the high frequency Lorentzian versus the potential for the SET island. The potential is calculated from the orthodox theory for the bias and gate voltage at which the spectrum was recorded. Stars (red) represent measurements taken at positive gain (D points), whereas circles (blue) represent measurements taken at negative slopes (B points). The line is a fit to eq.\,\ref{Eq_Cut_off frequency}.}
\end{figure}
Inserting the calculated potential into eq. (\ref{Eq_Cut_off frequency}) we can thus make a quantitative comparison between our model and the measured data.
In figure\,\ref{Fit} we plot the cut-off frequency of the high frequency Lorentzian as a function of the SET island potential.  As can be seen there is a good agreement between the data and our model, (eq.\,\ref{Eq_Cut_off frequency}). To obtain this we have used three fitting parameters, namely $\mathcal{A}$, $\mathcal{B}$ and $\phi_0$ of equation (\ref{Eq_Cut_off frequency}). From this fit we can extract important parameters of the grain SEB. 
From the $\mathcal{A}$ parameter we get the tunnel resistance to the grain, using the capacitance ratio extracted from the integrated charge change of the Lorentzian $\Delta q_{hf}$. The extracted tunnel resistance is $R'= 2.4\,\rm{G\Omega}$. Considering the size of the grain and that the sample was exposed to air before the measurement, this is quite reasonable. From the $\mathcal{B}$ parameter we can extract the temperature of the electrons in the lead, $T_{lead}=130\,\rm{mK}$. The experiment is performed at a mixing chamber temperature of $25\,\rm{mK}$, however we also have to take into account that SETs are always substantially overheated above the cryostat base temperature \cite{JApplPhys.78.2830}. Typically the overheating of an SET island is a few hundred mK. The lead next to the island will be colder, which is consistent with the temperature we extract. We can also see from the fit that the potential of the SET island passes through $\phi_0=0.22$, where the grain is at its degeneracy point. 

From the above discussion we can now also consider the low frequency Lorentzian and try to understand why its cut-off frequency is not changing with bias or gate voltage. As can be see in figure\,\ref{Gate_Depend}(c and d) we do not observe any significant change with the island potential. Assuming that the low frequency Lorentzian is also due to a grain, then this fluctuator must be far from its grain degeneracy point and therefore the relative frequency change is very small.  In this case the tunnel resistance to this grain must be substantially higher than that of the grain responsible for the high frequency Lorentzian.

If we consider a large ensemble of many grains close to the tunnel junctions of an SET, most of them will only be weakly coupled to the island and will thus together make up a $1/f$ background  in the noise spectrum. A few of the grains may be more strongly coupled to the island and will show up as individual Lorentzians in the noise spectrum just as we and many others have observed. In general it will be relatively rare that a strongly coupled grain is close to its degeneracy point since the charging energy of the grain is much larger than that of the SET. In addition we do need a large charging energy of the SET in order to change the frequency of the fluctuator. Thus it will be quite rare that a Lorentzian will show the large change in the cut-off frequency which we have observed. Therefore it is not so surprising that we have not been able to find this type of behavior in more samples in spite of numerous efforts.

The results presented here does not exclude that there are also other types of fluctuators with different physical mechanisms that contribute to the noise, however we are confident that  this is one type of fluctuator which we have been able to identify and found a good way to characterize.

\section{Conclusion}
In conclusion we have measured the noise of a single electron
transistor from a few Hz up to $10\,\mathrm{MHz}$. We find a spectrum with two Lorentzians superimposed on a $1/f$ background. The cut-off frequency of one of the Lorentzians depends strongly on the bias and gate voltages whereas the other does not. Our data is consistent with a model where the low-frequency noise comes from the random charge process of two effective single electron boxes coupled to the SET. We suggest that these single electron boxes are due to small aluminum grains coupled by tunneling to one of the leads and capacitively to the SET island. We are able to fit our data to this model with good agreement and we can extract parameters for the one of the fluctuators. 

\begin{acknowledgments}
The samples were fabricated in the MC2 clean room. This work was supported by the Swedish VR and SSF, and by the Wallenberg foundation.
\end{acknowledgments}

\bibliographystyle{plain}
\bibliography{References}

\end{document}